\documentclass[11pt, a4paper, logo, copyright, nonumbering]{deepseek}
\usepackage[authoryear, sort&compress, round]{natbib}
\usepackage{dblfloatfix}
\usepackage{ulem}
\usepackage{caption}
\usepackage{dramatist}
\usepackage{xspace}
\usepackage{pifont}
\usepackage{multirow}
\usepackage{tcolorbox}
\usepackage{adjustbox}
\usepackage{xltabular}
\usepackage{longtable}
\usepackage{hyperref}
\usepackage{amsfonts}
\usepackage{amsmath}
\usepackage{amssymb}
\usepackage{lineno}
\usepackage{algorithm}
\usepackage{algpseudocode}
\usepackage[bottom]{footmisc}
\usepackage{CJKutf8}
\usepackage{subfigure}
\usepackage{setspace}

\makeatletter
\def\@BTrule[#1]{%
  \ifx\longtable\undefined
    \let\@BTswitch\@BTnormal
  \else\ifx\hline\LT@hline
    \nobreak
    \let\@BTswitch\@BLTrule
  \else
     \let\@BTswitch\@BTnormal
  \fi\fi
  \global\@thisrulewidth=#1\relax
  \ifnum\@thisruleclass=\tw@\vskip\@aboverulesep\else
  \ifnum\@lastruleclass=\z@\vskip\@aboverulesep\else
  \ifnum\@lastruleclass=\@ne\vskip\doublerulesep\fi\fi\fi
  \@BTswitch}
\makeatother

\addto\extrasenglish{
}

 {\begin{list}{}%
         {\setlength{\leftmargin}{#1}}%
         \item[]%
 }
 {\end{list}}
 
\reportnumber{001} 

\renewcommand{\today}{}

\title{\centering 
DeepSeek-Coder-V2: Breaking the Barrier of Closed-Source Models in Code Intelligence
}
\author[*]{

\small
Qihao Zhu*, Daya Guo*, Zhihong Shao*, Dejian Yang*, Peiyi Wang, Runxin Xu, Y. Wu\newline
Yukun Li, Huazuo Gao, Shirong Ma, Wangding Zeng, Xiao Bi, Zihui Gu, Hanwei Xu, Damai Dai\newline
Kai Dong, Liyue Zhang, Yishi Piao, Zhibin Gou, Zhenda Xie, Zhewen Hao, Bingxuan Wang\newline 
Junxiao Song, Deli Chen, Xin Xie, Kang Guan, Yuxiang You, Aixin Liu, Qiushi Du, Wenjun Gao \newline
 Xuan Lu, Qinyu Chen, Yaohui Wang, Chengqi Deng, Jiashi Li, Chenggang Zhao \newline
Chong Ruan, Fuli Luo, Wenfeng Liang

\small
DeepSeek-AI \\
\small
\url{https://github.com/deepseek-ai/DeepSeek-Coder-V2}
\vspace{-0.2in}
}
\correspondingauthor{Core contributors
}

\renewcommand{\phi}{\varphi}

\renewcommand{\epsilon}{\varepsilon}
\renewcommand{\imath}{\mathrm{i}}

\newlength{\restsubwidth}
\newlength{\restsubheight}
\newlength{\restsubmoreheight}
\newcommand{\rest}[2]{%
        \settowidth{\restsubwidth}{\ensuremath{#2}}
        \settoheight{\restsubheight}{\ensuremath{{}_{#2}}}
        \ensuremath{{#1\hskip 0.5pt}_{\vrule\kern2pt\parbox[b][%
        4pt][b]{\the\restsubwidth}{%
                        \ensuremath{{}_{#2}}}}}
        }

\newcommand{\dscoder}{DeepSeek-Coder-V2\xspace}

\newcommand{\shortdsinslite}{DS-Coder-V2-Lite-Instruct\xspace}
\newcommand{\shortdsins}{DS-Coder-V2-Instruct\xspace}

\begin{abstract}
\vspace{-0.1in}
We present \dscoder,  an open-source Mixture-of-Experts (MoE) code language model that achieves performance comparable to GPT4-Turbo in code-specific tasks. Specifically, \dscoder is further pre-trained from an intermediate checkpoint of DeepSeek-V2 with additional 6 trillion tokens. Through this continued pre-training, \dscoder substantially enhances the coding and mathematical reasoning capabilities of DeepSeek-V2, while maintaining comparable performance in general language tasks. Compared to  DeepSeek-Coder-33B, \dscoder demonstrates significant advancements in various aspects of code-related tasks, as well as reasoning and general capabilities.  Additionally, \dscoder expands its support for programming languages from 86 to 338, while extending the context length from 16K to 128K.
In standard benchmark evaluations, \dscoder achieves superior performance compared to closed-source models such as GPT4-Turbo, Claude 3 Opus, and Gemini 1.5 Pro in coding and math benchmarks.  
\end{abstract}

\begin{document}

\begin{CJK*}{UTF8}{gbsn}

\maketitle



\begin{figure}[h]
    \centering
    \includegraphics[width=0.92\linewidth]{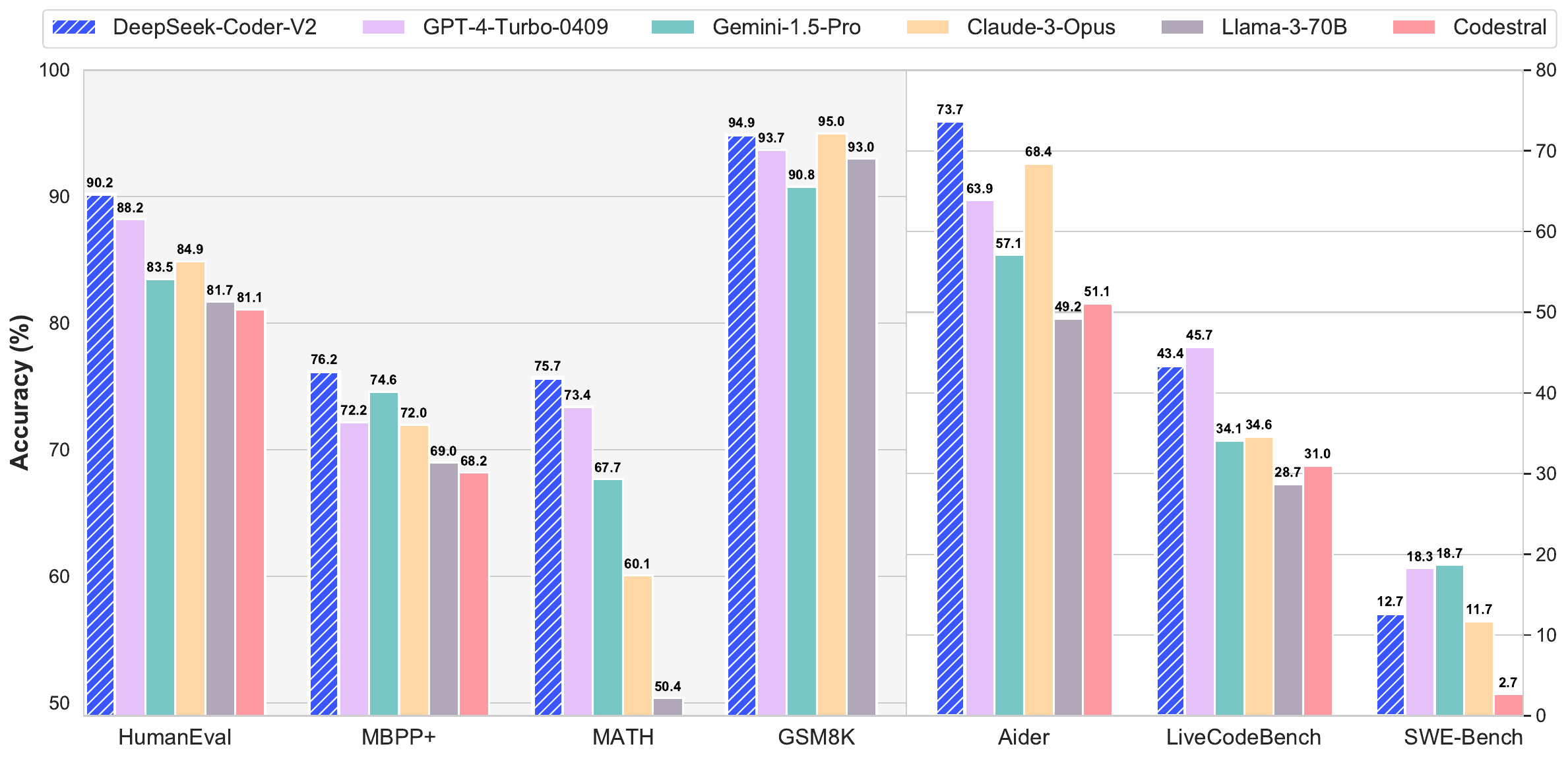}
    \caption{\centering The Performance of \dscoder on math and code benchmarks.}
\label{fig:data}
\end{figure}
\newpage

\section{Introduction}
\label{par:intro}
The open-source community has made significant strides in advancing code intelligence through the development of open-source code models such as StarCoder \citep{li2023starcoder,lozhkov2024starcoder}, CodeLlama \citep{roziere2023code}, DeepSeek-Coder \citep{guo2024deepseek}, and Codestral \citep{mistral2024codestral}. These models have steadily approached the performance levels of closed-source counterparts, contributing to the progress of code intelligence. However, there remains a discernible gap when comparing them to state-of-the-art closed-source models like GPT4-Turbo \citep{openai2023gpt4}, Claude 3 Opus \citep{anthropic2024claude}, and Gemini 1.5 Pro \citep{reid2024gemini}. To bridge this gap and further propel the development of open-source code models, we introduce the \dscoder series. These models are built upon the foundation of DeepSeek-V2 \citep{deepseekai2024deepseekv2} and are further pre-trained with an additional corpus with 6 trillion tokens.

In the pre-training phase, the dataset of \dscoder is created with a composition of 60\% source code, 10\% math corpus, and 30\% natural language corpus. The source code consists of 1,170B code-related tokens sourced from GitHub and CommonCrawl, using the same pipeline as DeepSeekMath \citep{shao2024deepseekmath}. This corpus expands from 86 to 338 programming languages compared to the code corpus used to train DeepSeek-Coder. To demonstrate the effectiveness of the new code corpus, we conduct ablation studies with the 1B parameter model and observe improvements of 6.7\% and 9.4\% in accuracy across both HumanEval (from 30.5\% to 37.2\%) and MBPP (from 44.6\% to 54.0\%) benchmarks ~\citep{chen2021evaluating,austin2021program}, respectively.  For the math corpus, we collect 221B math-related tokens sourced from CommonCrawl using the same pipeline, which approximately doubles the size of the 120B DeepSeekMath corpus \citep{shao2024deepseekmath}, while for the natural language corpus, we directly sample from the training corpus in DeepSeek-V2. In total, \dscoder has been exposed to 10.2T training tokens, where 4.2 trillion tokens originate from the  DeepSeek V2 dataset, while the remaining 6 trillion tokens come from the \dscoder dataset.

To accommodate longer code inputs and enhance applicability across various programming scenarios, we extend the context length from 16K to 128K tokens, allowing our models to handle more complex and extensive coding tasks. After continuous pre-training DeepSeek-V2  on this multi-source corpora, we find that \dscoder significantly enhances the model's capabilities in coding and mathematical reasoning while maintaining comparable general language performance. 

In the alignment phase, we first construct an instruction training dataset that includes code and math data from DeepSeek-Coder \citep{guo2024deepseek} and DeepSeek-Math \citep{shao2024deepseekmath}, as well as general instruction data from DeepSeek-V2 \citep{deepseekai2024deepseekv2}. This dataset is used to fine-tune the base model. Then, in the reinforcement learning phase, we employ Group Relative Policy Optimization (GRPO) algorithm to align its behavior with human preferences. Preference data is collected in the coding domain using compiler feedback and test cases, and a reward model is developed to guide the training of the policy model. This approach ensures that the model's responses are optimized for correctness and human preference in coding tasks. To enable the model to support code completion after alignment, we also utilize Fill-In-Middle approach \citep{guo2024deepseek} during the fine-tuning of the base model with 16B  parameters.

\subsection{Contributions}
In summary, our main contributions are:
\begin{itemize}[leftmargin=20pt]
    \item We introduce \dscoder with 16B and 236B parameters based on the DeepSeekMoE framework, which has activation parameters of only 2.4B and 21B, efficiently supporting diverse computational and application needs. Additionally, \dscoder supports 338 programming languages and a maximum context length of 128K tokens.
    \item  We make the first attempt to develop an open-source hundred-billion-parameter code model to advance the field of code intelligence. Experimental results indicate that \dscoder 236B outperforms state-of-the-art closed-source models, such as GPT4-Turbo, Claude 3 Opus, and Gemini 1.5 Pro, in both coding and mathematics tasks.
    \item \dscoder models are released publicly under a permissive license, allowing for both research and unrestricted commercial use.
\end{itemize}

\subsection{Summary of Evaluations and Metrics}
\begin{itemize}[topsep=0pt]
      \item \textbf{Code}: Regarding code generation benchmark evaluation, \dscoder demonstrates remarkable superiority over all open source models while exhibiting performance on par with the leading closed-source models, such as GPT4-Turbo, Claude 3 Opus, and Gemini 1.5 Pro. Notably, we achieve a \textbf{90.2}\% score on HumanEval \citep{chen2021evaluating}, a \textbf{76.2}\% score on MBPP \citep{austin2021program} (establishing a new state-of-the-art result with EvalPlus evaluation pipeline), and a \textbf{43.4}\% score on LiveCodeBench \citep{jain2024livecodebench} (questions from Dec. 2023 to June. 2024). Additionally, \dscoder is the first open-source model that  surpasses a score of 10\% on SWEBench   \citep{jimenez2023swe}. 

    \item \textbf{Math}:
    \dscoder exhibits strong mathematical reasoning abilities, rivaling top closed-source models such as GPT-4o, Gemini 1.5 Pro, and Claude 3 Opus on both elementary benchmarks like GSM8K \citep{gsm8k} and advanced competition-level benchmarks including MATH \citep{MATH}, AIME \citep{AIME}, and Math Odyssey \citep{netmindmath}.
    Notably, \dscoder attains an accuracy of \textbf{75.7}\% on the MATH benchmark, nearly matching the state-of-the-art accuracy of \textbf{76.6}\% achieved by GPT-4o.
    Furthermore, it surpasses the performance of these closed-source models in the AIME 2024 competition.

    

    \item \textbf{Natural Language}:  \dscoder maintains comparable general language performance to DeepSeek-V2. For example, \dscoder achieves 79.2\% on MMLU with OpenAI simple-eval pipeline. Regarding subjective evaluation with GPT-4 as a judger, \dscoder achieves \textbf{65.0} on arena-hard \citep{arenahard2024}, \textbf{8.77} on MT-bench \citep{mtbench} and  \textbf{7.84} on alignbench \citep{align_bench}. These scores are significantly better than other code-specific models, even comparable with general open source models.
\end{itemize}

\section{Data Collection}
The pre-training data for \dscoder primarily consists of 60\% source code, 10\% math corpus, and 30\% natural language corpus. Since the natural language corpus is directly sampled from the training dataset of DeepSeek-V2, this section focuses on the collection, cleaning, and filtering processes of the code and math data. Meanwhile, we further validate the quality of this data through comparative analysis experiments.

We collect public repositories created before November 2023 on GitHub. We first apply the same filtering rules and near-deduplication as those used in the DeepSeek-Coder \citep{guo2024deepseek} to filter out lower-quality and duplicated source code. To make the paper self-contained, we briefly describe the filtering rules. Firstly, we filter out files with an average line length exceeding 100 characters or a maximum line length surpassing 1000 characters. Additionally, we remove files with fewer than 25\% alphabetic characters. Except for the XSLT programming language, we further filter out files where the string \texttt{"<?xml version="} appears in the first 100 characters. For HTML files, we consider the ratio of visible text to HTML code. We retain files where the visible text constitutes at least 20\% of the code and is no less than 100 characters. For JSON and YAML files, which typically contain more data, we only keep files that have a character count ranging from 50 to 5000 characters. This effectively removes most data-heavy files. By applying these filtering rules and near-deduplication, we obtain 821B  code encompassing 338 programming languages and 185B code-related text, such as markdown and issues. The list of supported programming languages can be found in the Appendix \ref{sec:supported_pl}. We use the same tokenizer as DeepSeekV2,  detailed in \citep{deepseekai2024deepseekv2}.

To collect code-related and math-related web texts from Common Crawl, we follow the same pipeline as DeepSeekMath \citep{shao2024deepseekmath}. Specifically, we select coding forums such as StackOverflow\footnote{https://stackoverflow.com}, library sites such as PyTorch documentation\footnote{https://pytorch.org/docs}, and mathematics website such as StackExchange\footnote{https://math.stackexchange.com} 
as our initial seed corpus. Using this seed corpus, we train a fastText model \citep{joulin2016fasttext} to recall more coding-related and math-related web pages. Since tokenization for languages like Chinese cannot be done through spaces, we use the Byte Pair Encoding (BPE) tokenizer from DeepSeek-V2, which significantly improves the recall accuracy of fastText. For each domain, we calculate the percentage of web pages collected in the first iteration. Domains with over 10\% of web pages collected are classified as code-related or math-related. We then annotate the URLs associated with code-related or math-related content within these identified domains. Uncollected web pages linked to these URLs are added to the seed corpus. After three iterations of data collection, we gather 70 billion code-related tokens and 221B math-related tokens from web pages.
To further collect high-quality source code from GitHub, we also apply the same pipeline on GitHub with two iterations of data collection and collect 94B source code. The initial seed corpus is constructed by manually collecting high-quality source code, such as those containing detailed descriptions. Finally, the new code corpus consists of 1,170B code-related tokens sourced from GitHub and CommonCrawl.

To demonstrate the effectiveness of the new code corpus, we conducted ablation studies (see Table \ref{table:code-corpus-ablation}) using a 1B parameter model, comparing it with the corpus used to train DeepSeek-Coder. Pre-training the 1B model on the new code corpus using 1T tokens resulted in improvements of 5.5\% and 4.4\% in accuracy on the HumanEval (from 30.5\% to 36.0\%) and MBPP (from 44.6\% to 49.0\%) benchmarks, respectively. Further training the 1B model with 2T tokens led to additional improvements, with HumanEval and MBPP scores rising to 37.2\% and 54.0\%, respectively. Therefore, the new code corpus is superior to the code corpus used to train DeepSeek-Coder.
\begin{table}[h]

        \begin{small}
	
		\centering
		\resizebox{\linewidth}{!}{\begin{tabular}{lc|c|c|c|c|c|c|c|c|c|c}
			\toprule
			Model& Tokens&Python&C++&Java&PHP&TS&C\#&Bash&JS&Avg &MBPP \\

               \midrule
                DeepSeek-Coder-1B &1T&30.5\%&28.0\%&31.7\%&23.0\%&30.8\%&31.7\%&9.5\%&28.6\%&26.7\% &44.6\% \\
                \dscoder-1B&1T&36.0\%&34.8\%&31.7\%&27.3\%&\bf{37.7\%}&34.2\%&6.3\%&\bf{38.5\%}&31.2\% &49.0\% \\                
                \dscoder-1B &2T&\bf{37.2\%}&\bf{39.1\%}&\bf{32.3\%}&\bf{31.7\%}&34.6\%&\bf{36.7\%}&\bf{12.0\%}&32.9\%&\bf{32.0\%} &\bf{54.0\%}\\
                \bottomrule
		\end{tabular}}
	\caption{\centering Performance of 1B base model between DeepSeek-Coder and \dscoder.}
        \label{table:code-corpus-ablation}
        \end{small}
\vspace{-0.1in}
\end{table}

\section{Training Policy}
\subsection{Training Strategy}
We use two training objectives for DeepSeek-Coder-v2 16B: Next-Token-Prediction and Fill-In-Middle (FIM)  ~\citep{li2023starcoder,bavarian2022efficient,guo2024deepseek}. For DeepSeek-Coder-v2 236B, we only utilize the Next-Token-Prediction objective.
Here we give a brief introduction of the FIM training policy. We adopt the FIM training approach for the development of DeepSeek-Coder-v2-16B, leveraging the PSM (Prefix, Suffix, Middle) mode. This method structures the content reconstruction in the sequence: Prefix, Suffix, and Middle, as illustrated below:

\begin{align}
\texttt{<｜fim\_begin｜>}f_{pre}\texttt{<｜fim\_hole｜>}f_{suf}\texttt{<｜fim\_end｜>}f_{middle}\texttt{<|eos\_token|>} \nonumber
\end{align}
This structure is applied at the document level as part of the pre-packing process. The FIM is utilized at a rate of 0.5, consistent with the PSM framework, to enhance the training efficacy and model performance.

\subsection{Model Architecture}
Our architecture aligns with that of DeepSeekV2 \citep{deepseekai2024deepseekv2}. The hyperparameters settings, 16B and 236B, correspond to those used in DeepSeek-V2-Lite and DeepSeek-V2, respectively. Notably, we encountered instability during training and spikes in gradient values, which we attributed to the exponential normalization technique. To address this, we reverted to the conventional normalization method.
\subsection{Training Hyper-Parameters} 

Consistent with the DeepSeek V2 methodology \citep{deepseekai2024deepseekv2}, we utilize the AdamW optimizer \citep{loshchilov2019decoupled}, configured with $\beta_1 = 0.9$, $\beta_2 = 0.95$, and a weight decay of 0.1. Batch sizes and learning rates are adjusted according to DeepSeek-V2 specifications. For learning rate scheduling, we employ a cosine decay strategy, starting with 2000 warm-up steps and gradually reducing the learning rate to 10\% of its initial value.

Both \dscoder and \dscoder-Lite are trained using the same methodology. To maintain robust natural language understanding capabilities in \dscoder, we continue the pre-training process from an intermediate checkpoint of DeepSeek-V2. The intermediate checkpoint was initially trained on 4.2T tokens. Consequently, \dscoder has been exposed to a total of 10.2T high-quality tokens during the pre-training phase.
\begin{table}[h]

  \centering
	       \begin{small}
	           
		\begin{tabular}{l|c|c}
			\toprule
			Model& \dscoder-Lite&\dscoder \\ \hline
                \# Total Parameters (\#TP) & 16B & 236B \\ \hline
                \# Active Parameters (\#AP) & 2.4B & 21B \\ \hline
                Pre-training Tokens& 4.2T+6T &  4.2T+6T \\ \hline
                LR Scheduler& Cosine & Cosine \\ \hline
                FIM & Enable & Disable \\

                \bottomrule
		\end{tabular}
  \end{small}
	\caption{\centering Training Setting of \dscoder. }
\vspace{-0.2in}
\end{table}
\subsection{Long Context Extension}
\begin{figure}[!t]
\centering
\includegraphics[width=0.82\linewidth]{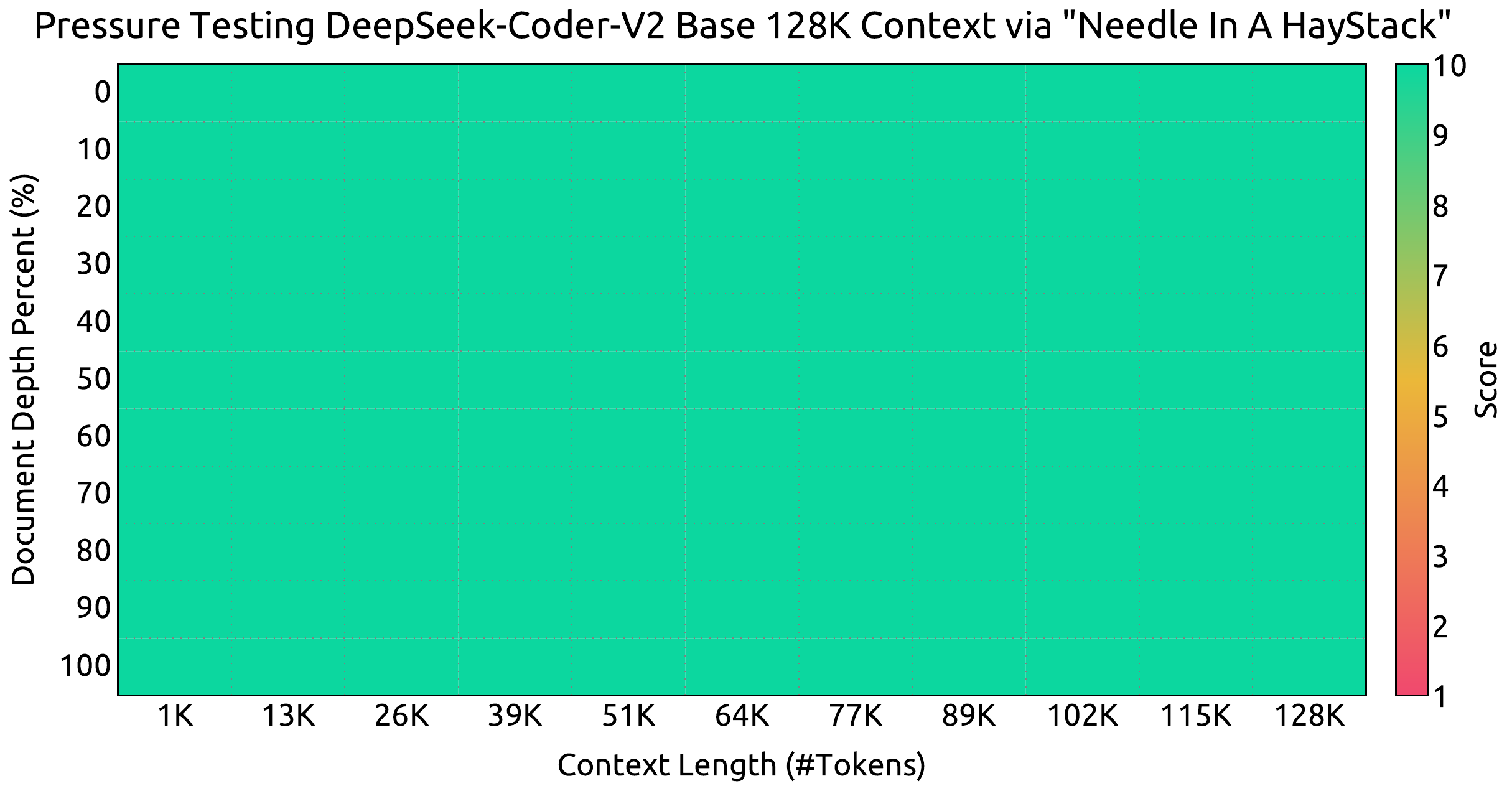}
\caption{
\centering Evaluation results on the ``Needle In A Haystack'' (NIAH) tests. DeepSeek-Coder-V2 performs well across all context window lengths up to 128K. 
}
\label{fig:long_context}
\vspace{-0.1in}
\end{figure}

Following DeepSeek-V2, we extend the context length of DeepSeek-Coder-V2 to 128K using Yarn \citep{peng2023yarn}. The hyper-parameters of YARN are the same as DeepSeek-V2: the scale $s$ to 40, $\alpha$ to 1, $\beta$ to 32.
 We further continue training the model using two stages to enhance its capability for handling long contexts. In the first stage, we utilize a sequence length of 32K and a batch size of 1152 for 1000 steps. In the second stage, we train the model for an additional 1000 steps, employing a sequence length of 128K and a batch size of 288 sequences. It should be noted here we upsample long context data ratio during long context extension.  As shown in Figure~\ref{fig:long_context}, the results on the ``Needle In A Haystack'' (NIAH) tests indicate that   DeepSeek-Coder-V2  performs well across all context window lengths up to 128K.

\subsection{Alignment}
\subsubsection{Supervised Fine-Tuning} To build DeepSeek-Coder-V2 Chat, we construct the instruction training dataset mixed with code and math data. We first collect 20k code-related instruction data and 30k math related data from DeepSeek-Coder and DeepSeek-Math. To maintain the general ability, we also sample several data from the instruction data of DeepSeek-V2. Finally, we use a instruction dataset of 300M tokens. For training, we use a cosine schedule with 100 warm-up steps and an initial learning rate $5e^{-6}$. We also use a batch size of 1M tokens and 1B tokens in total. 

\subsubsection{Reinforcement Learning}
We further employ Reinforcement Learning (RL) techniques to fully simulate the capabilities of \dscoder, which is proven to be quite effective.

\paragraph{Prompts}
Considerable effort was spent collecting prompts related to code and math from various sources, and each code prompt comes with corresponding test cases.
After filtering the prompts, there are approximately 40k data in total.

\begin{figure}[h]
    \centering
    \includegraphics[width=0.92\linewidth]{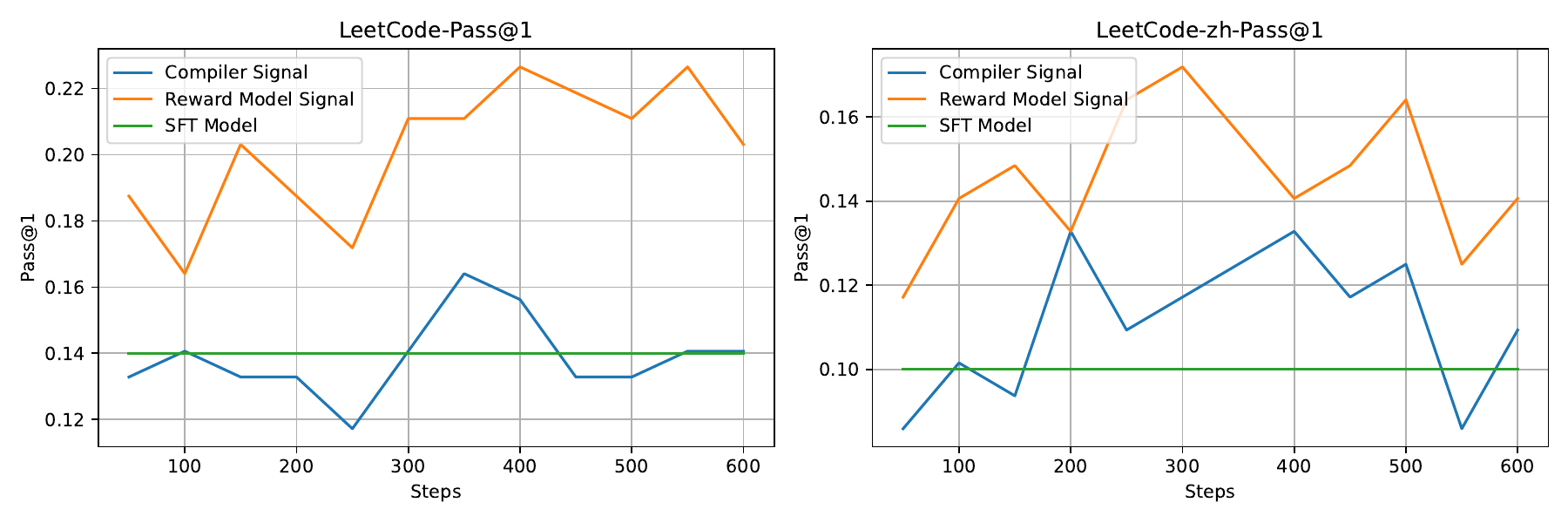}
    \caption{\centering Performances of Different Methods}
\label{fig:rl}

\end{figure}

\paragraph{Reward Modeling}
Reward models play crucial roles in the RL training.
In terms of mathematical preference data, we obtain them using the ground-truth labels. 
In terms of code preference data, although the code compiler itself can already provide 0-1 feedback (whether the code pass all test cases or not), some code prompts may have a limited number of test cases, and do not provide full coverage, and hence directly using 0-1 feedback from the compiler may be noisy and sub-optimal.
Therefore, we still decide to train a reward model on the data provided by the compiler, and use the reward model to provide signal during RL training, which is more robust and has better generalization ability, in comparison with raw compiler signal.
As illustrated in Figure \ref{fig:rl}, in our in-house test sets (Leetcode and Leetcode-zh), using a reward model to provide RL training signal clearly outperforms using raw compiler signal. 
Hence, we use reward model signal rather than compiler signal in all subsequent experiments.

\paragraph{Reinforcement Learning Algorithm}
We employ Group Relative Policy Optimization (GRPO) \cite{shao2024deepseekmath} as our RL algorithm, which is the same as what DeepSeek-V2 uses. 
Notably, GRPO is proven to be quite effective and has less cost compared with PPO, since there is no need to maintain an additional critic model.



\section{Experimental Results}
In this section, we evaluate \dscoder on three types of tasks, including coding, mathematics, and general natural language. We compare \dscoder with the previous state-of-the-art large language models.
\begin{itemize}
    \item \textbf{CodeLlama} \citep{roziere2023code} consists of a series of code language models  based on Llama2 \citep{touvron2023llama}, and continue pre-training on datasets ranging from 500 to 1000 billion code tokens. These models are available in four sizes: 7B, 13B, 34B, and 70B. 
    \item \textbf{StarCoder} \citep{lozhkov2024starcoder} is a publicly accessible model with 15 billion parameters. It is specifically trained on a meticulously curated subset of the Stack dataset \citep{kocetkov2022stack}, covering 86 programming languages.
    
    \item \textbf{StarCoder2} \citep{lozhkov2024starcoder} consists of 3B, 7B, and 15B parameters models trained on 3.3 to 4.3 trillion tokens of the Stack2 dataset \citep{lozhkov2024starcoder}, spanning 619 programming languages.

    \item \textbf{DeepSeek-Coder} \citep{guo2024deepseek} comprises a series of code language models, ranging from 1 billion to 33 billion parameters. Each model is trained from scratch on 2 trillion tokens, with a composition of 87\% code and 13\% natural language in both English and Chinese. These models are pre-trained on a project-level code corpus using a window size of 16K and an additional fill-in-the-blank task, enabling support for project-level code completion and infilling.
    
    \item \textbf{Codestral} \citep{mistral2024codestral} is a 22B parameter model developed by Mistral. It is trained on a diverse dataset of over 80 programming languages, including popular ones such as Python, Java, and JavaScript, as well as more specialized languages like Swift and Fortran.
    
    \item General language models that we compare include \textbf{Llama3 70B} \citep{meta2024llama3}, \textbf{GPT-4} \citep{openai2023gpt4}, \textbf{Claude 3 Opus} \citep{anthropic2024claude}, and \textbf{Gemini 1.5 Pro} \citep{reid2024gemini}. While they are not specifically trained on large code corpora, they achieve state-of-the-art performance in coding.
\end{itemize}
\subsection{Code Generation} 
\paragraph{HumanEval and MBPP Benchmarks.}The HumanEval \citep{chen2021evaluating} \footnote{We use the template "Please complete the python function below. The final complete version of your function must be returned within a code block. Here is the unfinished function:\textbackslash n \textasciigrave\textasciigrave\textasciigrave python\textbackslash n\{problem\_description\}\textbackslash n\textbackslash n" to build the instruction prompt.} and MBPP \citep{mbpp} benchmarks are commonly utilized for assessing the performance of code-generating Large Language Models (LLMs). HumanEval comprises 164 Python tasks that are verified through test cases to evaluate the performance of Code LLMs in a zero-shot scenario. For MBPP, we use the MBPP-Plus version \citep{evalplus} to evaluate the models. To test the multilingual abilities of models, we extended the HumanEval benchmark problems into seven additional languages: C++, Java, PHP, TypeScript, C\#, Bash, JavaScript, Swift, R, Julia, D, Rust and Racket. For both benchmarks, we employed a greedy search strategy and recreated the baseline results using identical scripts and environments to ensure a fair comparison. 
\begin{table}[h!]
\begin{small}
\centering
\begin{adjustbox}{max width=\textwidth}
\begin{tabular}{lcccccccccccccccccc}
\toprule
&\#TP&\#AP& Python & Java & C++ & C\# & TS & JS & PHP & Bash \\
\midrule
  \multicolumn{11}{c}{Closed-Source Models} \\ \midrule
Gemini-1.5-Pro&-&- & 83.5\% & 81.0\% & 78.3\% & 75.3\% & 77.4\% & 80.8\% & 74.5\% & 39.9\% \\
Claude-3-Opus&-&- & 84.2\% & 78.5\% & 81.4\% & 74.7\% & 76.1\% & 75.8\% & 78.3\% & 48.7\%  \\
GPT-4-1106&-&- & 87.8\% & \textbf{82.3\%} & 78.9\% & 80.4\% & 81.8\% & 80.1\% & 77.6\% & \textbf{55.7\%} \\
GPT-4-Turbo-0409&-&- & 88.2\% & 81.7\% & 78.3\% & 79.1\% & 79.3\% & 80.8\% & 78.9\% & \textbf{55.1\%} \\
GPT-4o-0513&-&- & \textbf{91.0\%} & 80.4\% & \textbf{87.0\%} & \textbf{82.9\%} & \textbf{86.2\%} & \textbf{87.6\%} & \textbf{79.5\%} & 53.8\% \\
\midrule
  \multicolumn{11}{c}{Open-Source Models} \\ \midrule
Codestral&22B&22B & 78.1\% & 71.5\% & 71.4\% & 77.2\% & 72.3\% & 73.9\% & 69.6\% & 47.5\% \\
DS-Coder-instruct &33B&33B&{79.3\%}&73.4\%&68.9\% &74.1\%&67.9\%&73.9\%&72.7\%&43.0\%\\

Llama3-Instruct&70B&70B&81.1\%&67.7\%&64.0\%&69.6\%&69.8\%&70.2\%&65.8\%&36.1\%\\
\midrule
\shortdsinslite &16B&2.4B&81.1\%&76.6\%&75.8\%&76.6\%&80.5\%&77.6\%&74.5\%&43.0\%\\
\shortdsins &236B&21B&\textbf{90.2\%}&\textbf{82.3\%}&\textbf{84.8\%}&\textbf{82.3\%}&\textbf{83.0\%}&\textbf{84.5\%}&\textbf{79.5\%}&\textbf{52.5\%}\\

\bottomrule
\toprule
&\#TP&\#AP& Swift & R & Julia & D & Rust & Racket&$\text{MBPP}^{+}$ & Average  \\\midrule
  \multicolumn{11}{c}{Closed-Source Models} \\ \midrule
Gemini-1.5-Pro&-&- & 66.5\% & 53.4\% & 71.7\% & 55.8\%  & 73.1\% & 48.4\% & \textbf{74.6\%}&68.9\% \\
Claude-3-Opus&-&- & 63.9\% & 55.9\% & 76.1\% & 60.3\%  & 71.2\% & \textbf{64.6\%} & 72.0\% &70.8\%\\
GPT-4-1106&-&- & 62.7\% & 57.8\% & 69.2\% & 60.9\%  & \textbf{78.8\%} & 64.0\% & 69.3\%&72.5\% \\
GPT-4-Turbo-0409&-&- & 63.9\% & 56.5\% & 69.8\% & \textbf{61.5\%} & \textbf{78.8\%} & 63.4\% & 72.2\%&72.3\% \\
GPT-4o-0513 &-&-& \textbf{75.9\%} & \textbf{65.2\%} & \textbf{78.0\%} & 60.9\% & \textbf{80.1\%} & \textbf{64.6\%} & {73.5\%}&\textbf{76.4\%} \\

\midrule
  \multicolumn{11}{c}{Open-Source Models} \\ \midrule
Codestral&22B&22B & 63.3\% & 49.7\% & 67.9\% & 32.1\%  & 67.3\% & 37.3\% & 68.2\%&63.2\% \\
DS-Coder-instruct &33B&33B&{61.4\%}&44.7\%&53.5\% &31.4\%&68.6\%&46.0\%&70.1\%&61.9\%\\
Llama3-Instruct&70B&70B&55.1\%&46.0\%&62.9\%&48.1\%&58.3\%&46.0\%&68.8\%&60.6\%\\
\midrule
\shortdsinslite&16B&2.4B&64.6\%&47.8\%&67.3\%&45.5\%&62.2\%&41.6\%&68.8\%&65.6\%\\
\shortdsins&236B&21B&\textbf{72.2\%}&\textbf{64.0\%}&\textbf{72.3\%}&\textbf{64.1\%}&\textbf{78.2\%}&\textbf{63.4\%}&\textbf{76.2\%}&\textbf{75.3\%}\\
\bottomrule
\end{tabular}
\end{adjustbox}
\caption{\centering Performance Metrics for Various Models on HumanEval and MBPP Benchmarks}
\label{table:results}
\end{small}
\end{table}

Table \ref{table:results} provides an extensive overview of the performance metrics for various models across multiple programming languages on the HumanEval and $\text{MBPP}^{+}$ Benchmarks. The DeepSeek-Coder-V2-Instruct  demonstrates exceptional performance, securing the second-highest average score of 75.3\%. This performance is notable as it breaks the dominance typically seen from closed-source models, standing out as a leading open-source contender. It is surpassed only by GPT-4o, which leads with an average score of 76.4\%. DeepSeek-Coder-V2-Instruct shows top-tier results across a variety of languages, including the highest scores in Java and PHP, and strong performances in Python, C++, C\#, TypeScript, and JavaScript, underscoring its robustness and versatility in handling diverse coding challenges.

Furthermore, the DeepSeek-Coder-V2-Lite-Instruct  also performs impressively, surpassing the larger 33B model. With a considerable margin in average performance (65.6\% vs. 61.9\%), it highlights the effectiveness of the 16B model in delivering competitive results despite its smaller size. This underscores the model's efficiency and the advancements in model architecture and training methodologies that allow it to outperform larger counterparts.
\paragraph{Competitive Programming.}To further validate the model's capability in real-world competitive programming problems, we utilize the LiveCodeBench \citep{jain2024livecodebench} and USACO benchmark \citep{usaco} to estimate the effectiveness of DeepSeek-Coder-V2. LiveCodeBench is a meticulous and contamination-free assessment of Large Language Models (LLMs) for code generation, systematically gathering novel challenges over time from three prominent competitive programming platforms: LeetCode, AtCoder, and CodeForces. Since the cut-off of the training data is before November 2023, we use the subset (1201-0601) of Livecodebench. USACO benchmark contains 307 problems from the USA Computing
Olympiad, along with high-quality unit tests, reference code, and official
analyses for each problem.

\begin{table}[h]
        \centering

		\centering
            \begin{small}
            \resizebox{\linewidth}{!}{
		\begin{tabular}{lcc|cccc|c}
			\toprule
                \multirow{2}{*}{Model}& \multirow{2}{*}{\#TP} &\multirow{2}{*}{\#AP} & \multicolumn{4}{c|}{LiveCodeBench}&\multirow{2}{*}{USACO}\\
			&  & &  Easy (82) &  Medium (87)& Hard (57)&  Overall (226)& \\
               \midrule 
                 \multicolumn{8}{c}{Closed-Source Models} \\ \midrule
                Gemini-1.5-Pro &-&-&74.9\%&16.8\%&1.8\%&34.1\%&4.9\%\\
                Claude-3-Opus&-&-&77.2\%&16.7\%&0.7\%&34.6\%&7.8\%\\
                GPT-4-1106&-&-&78.4\%&20.2\%&3.5\%&37.1\%&11.1\%\\
                GPT-4-Turbo-0409 &-& - &{84.1\%}&\textbf{35.4\%}&\textbf{6.1\%}& \textbf{45.7\%}&12.3\%\\  
                GPT-4o-0513 &-& - &\textbf{87.4}\%& 27.5\%&4.9\%& {43.4\%}&\textbf{18.8\%}\\

               \midrule
                 \multicolumn{8}{c}{Open-Source Models} \\ \midrule
                 Codestral&22B&22B&{66.5\%}&17.7\%&0.2\% &31.0\%&4.6\% \\
               DS-Coder-instruct &33B&33B&{51.6\%}&9.7\%&0.4\% &22.5\%&4.2\%\\

Llama3-Instruct&70B&70B&62.4\%&14.4\%&2.1\%&28.7\%&3.3\%\\
                \midrule
                \shortdsinslite &16B&2.4B&58.5\%&8.0\%&0.0\%&24.3\%&6.5\%\\
                \shortdsins &236B&21B&\textbf{84.1\%}&\textbf{29.9\%}&\textbf{5.3\%}&\textbf{43.4\%}&\textbf{12.1\%}\\

                \bottomrule 
		\end{tabular}}
	\caption{\centering  Performance on the LiveCodeBench (LCB) and USACO benchmarks.}
 \label{table:result-leetcode}
        \end{small}
\end{table}

Table \ref{table:result-leetcode} showcases the performance of various language models on the two benchmarks. Notably, DeepSeek-Coder-V2-Instruct  delivers a standout performance, tying for the highest score among large models at 43.4\%, on par with GPT-4o. This exceptional result places it second overall, just behind GPT-4-Turbo-0409, which leads with an overall performance of 45.7\%. DeepSeek-Coder-V2-Instruct's impressive ability to handle complex coding challenges firmly establishes it as a top contender, closely trailing the leading GPT-4-Turbo variant.

\subsection{Code Completion}
\subsubsection{Repository-Level Code Completion Evaluation}
We use RepoBench \citep{liu2023repobench} to evaluate the capabilities of currently available open-source code models with sizes below 35B in repository-level code completion tasks. This dataset is constructed from a diverse set of real-world, open-sourced, permissively licensed repositories in two popular programming languages: Python and Java. Notably, the latest version (v1.1) of RepoBench sources its data from GitHub repositories created between October 6th and December 31st, 2023, while our pre-training data includes code created before November 2023. To ensure this dataset was not present in our pre-training data and avoid data leakage, we only use data from December 2023.

Our evaluation includes five context length levels—2k, 4k, 8k, 12k, and 16k tokens—across three settings: cross-file-first, cross-file-random, and in-file. We use greedy search for all models under evaluation. The models were constrained to generate a maximum of 64 new tokens per prompt, and the first non-empty and non-comment line of the output was selected as the prediction. The maximum token length for prompts was set to 15,800 by truncating excess cross-file context. We report the average exact match for the different context length levels.
\begin{table}[h]
\centering
\begin{small}
\resizebox{\linewidth}{!}{\begin{tabular}{l l l |c c c c c c| c c c c c c}
\toprule
\multirow{2}{*}{Model} &\multirow{2}{*}{\#TP}&\multirow{2}{*}{\#AP} &\multicolumn{6}{c|}{Python}&\multicolumn{6}{c}{Java}\\
\cmidrule(lr){4-9}\cmidrule(lr){10-15}&&&2k&4k&8k&12k&16k&Avg&2k&4k&8k&12k&16k&Avg \\
\midrule
StarCoder2-Base & 15B & 15B& 35.7\% & 36.7\% &34.6\%& 27.4\% & 25.1\% & 32.1\% & 46.2\% & 45.0\% &39.8\%& 30.5\% & 30.7\% & 38.7\%\\
CodeLlama-Base & 7B & 7B& 32.0\% & 34.4\% &35.3\%& 33.3\% & 32.2\% & 33.5\% & 43.1\% & 42.1\% &40.4\%& 37.0\% & 40.3\% & 40.6\%\\
CodeLlama-Base & 13B & 13B& 33.0\% & 36.5\% &37.0\%& 34.6\% & 35.0\% & 35.2\% &  43.5\% & 44.8\% &40.7\%& 38.6\% & 41.1\% & 41.8\%\\
CodeLlama-Base & 34B & 34B& 35.3\% & 37.5\% &39.5\%& 34.9\% & 35.6\% & 36.6\% &   45.9\% & 45.4\% &42.5\%& 41.0\% & 41.2\% & 43.3\%\\
DS-Coder-Base & 6.7B & 6.7B& 36.1\% & 37.5\% &38.2\%& 34.0\% & 35.0\% & 36.2\% &  46.8\% & 46.4\% &42.9\%& 38.8\% & 40.8\% & 43.3\%\\
DS-Coder-Base & 33B & 33B& {39.7\%} & {40.1\%} &40.0\%&36.9\% &38.5\% &{39.1\%} &  47.9\% &{47.7\%} &{43.3\%}& {40.9\%} & {43.6\%} & {44.8\%}\\
Codestral & 22B & 22B& \bf{42.1\%} & \bf{44.3\%} &\bf{46.6\%}& \bf{46.6\%} & \bf{51.5\%} & \bf{46.1\%} & {48.3\%} & \bf{47.8\%} &\bf{46.0\%}& \bf{42.2\%} & \bf{43.9\%} & \bf{45.7\%}\\
\midrule
DS-Coder-V2-Lite-Base & 16B & 2.4B&  38.3\% & 38.6\% &{40.6\%}& {38.3\%} & {38.7\%} & 38.9\% &  \bf{48.8\%} & 45.7\%&42.4\%& 38.1\% & 41.1\% & 43.3\%\\


\bottomrule
\end{tabular}}

\caption{\centering Performance of different models on December subset of RepoBench v1.1.}
\label{tab:code_completion_results}
\end{small}

\end{table}

As shown in Table \ref{tab:code_completion_results}, the results indicate that the DeepSeek-Coder-V2-Lite-Base model, despite having only 2.4 billion active parameters, achieves code completion capabilities in Python comparable to the DeepSeek-Coder-Base 33B model and in Java comparable to the DeepSeek-Coder-Base 7B model. Compared to CodeStral, the DeepSeek-Coder-V2-Lite-Base model has only one-tenth of the active parameters of CodeStral, resulting in lower performance in code completion tasks. However, we believe that the smaller number of active  parameters in DeepSeek-Coder-V2 makes it faster for code completion scenarios.

\subsubsection{Fill-in-the-Middle Code Completion}
\dscoder-Lite is trained with a unique approach that includes a 0.5 Fill-In-the-Middle (FIM) rate during their pre-training phase. This method allows the model to adeptly complete code by filling in blanks using the surrounding context, which includes both the preceding and following code segments. This ability is particularly advantageous for code completion tools. Several open-source models, such as SantaCoder \citep{allal2023santacoder}, StarCoder \citep{li2023starcoder}, and CodeLlama \citep{roziere2023code}, also leverage similar capabilities and have established high standards in the domain of code generation and completion.

To evaluate the performance of \dscoder models, we conducted a comparative analysis against leading models. The assessment was based on the Single-Line Infilling benchmarks, covering three different programming languages as described by \citet{allal2023santacoder}. The main metric for this evaluation was the line exact match accuracy\footnote{We use the first generated line rather than the whole generated chunk, thus the result is slightly different with DeepSeek-Coder.}.

\begin{table}[h]
		\centering
  \begin{small}
		\begin{tabular}{lcccccc}
			\toprule
			Model&\#TP&\#AP&python&java&javascript &Mean\\
               \midrule    
                StarCoder\footnote{StartCoder-2 has some problems with FIM, thus we still use StartCoder.}&16B&16B&71.5\%&82.3\%&83.0\% &80.2\%\\
                CodeLlama-Base&7B&7B&58.6\%&70.6\%&70.7\% &68.0\%\\
                CodeLlama-Base&13B&13B&{60.7}\%&74.3\%&78.5\%&73.1\%\\
                \midrule
                DS-Coder-Base&1B&1B&74.1\%&85.1\%&82.9\%& 81.8\%\\
                DS-Coder-Base&7B&7B&79.8\%&\textbf{89.6}\%&86.3\%&86.1\%\\
                DS-Coder-Base&33B&33B&\textbf{80.5\%}&88.4\%&{86.6}\% &\textbf{86.4}\%\\
                Codestral&22B&22B&77.2\%&83.2\%&{85.9}\% &{83.0}\%\\
                \midrule
                DS-Coder-V2-Lite-Base &16B&2.4B&{80.0\%}&89.1\%&\textbf{87.2\%}&\textbf{86.4\%}\\
                \bottomrule 
		\end{tabular}
	\caption{\centering Performance of different approaches on the FIM-Tasks.}        
	\label{table:compare-to-other-alg-fim}
        
            \end{small}
\end{table}
The table presents the performance of various coding models on FIM (Fill-in-the-Middle) tasks across three programming languages: Python, Java, and JavaScript, with a Mean score indicating overall effectiveness. Among the compared models, \dscoder-Lite-Base, with a configuration of 2.4B active parameters, achieves outstanding results. It scores 80.0\% in Python, 89.1\% in Java, and 87.2\% in JavaScript, leading to a top Mean score of 86.4\%. This demonstrates the superior effectiveness of \dscoder-Lite-Base, particularly in handling FIM tasks across different programming languages, achieving comparable performance with other bigger models in the evaluation.

\subsection{Code Fixing }
To evaluate the bug-fixing capabilities of the model, we used the Defects4J \footnote{https://github.com/rjust/defects4j}, SWE-bench \citep{jimenez2023swe}, and Aider \footnote{https://github.com/paul-gauthier/aider} datasets for testing. Defects4J is a widely used dataset in the field of software engineering, specifically designed for the purpose of evaluating and testing program repair techniques. It consists of a collection of real-world software bugs from various open-source projects, including but not limited to Apache Commons, JFreeChart, and Closure Compiler. Each bug in the dataset is accompanied by test suites that can be used to validate the effectiveness of program repair tools. Since the original bugs in Defec4J may need modify several files in the repository resulting in a long context, we collect 238 bugs that only need to modify one method from this benchmark. 

SWE-bench is a comprehensive benchmark designed to evaluate the performance of large language models in addressing real-world software issues sourced from GitHub. The benchmark presents a codebase alongside a specific issue, challenging the language model to generate a patch that effectively resolves the described problem. This rigorous evaluation framework ensures that the language model's ability to understand and fix real-world software issues is thoroughly tested, providing a clear measure of its practical utility and effectiveness in software development tasks.

Aider's code editing benchmark evaluates the LLM's ability to modify Python source files, completing 133 distinct coding tasks. This benchmark not only tests the LLM's coding skills but also checks its consistency in producing code edits according to the specifications in the prompt. For DeepSeek-Coder-V2 models, we use \textit{whole} format to evaluate.
\begin{table}[h]
		\centering
  \begin{small}
		\begin{tabular}{lccccc}
			\toprule
   
			Model&\#TP&\#AP&Defects4J&SWE-Bench&Aider\\
               \midrule    
                             \multicolumn{6}{c}{Closed-Source Models} \\ \midrule
                Gemini-1.5-Pro&-&-&18.6\%&19.3\%&57.1\%\\
                Claude-3-Opus&-&-&25.5\%&11.7\%&68.4\%\\
                GPT-4-1106&-&-&22.8\%&22.7\%&65.4\%\\
                GPT-4-Turbo-0409&-&-&24.3\%&18.3\%&63.9\%\\
                GPT-4o-0513&-&-& \textbf{26.1}\%&\textbf{26.7}\%&\textbf{72.9\%}\\
                \midrule
                     \multicolumn{6}{c}{Open-Source Models} \\ \midrule
                Codestral&22B&22B&17.8\%&2.7\%&51.1\%\\
                DS-Coder-Instruct&33B&33B&{11.3\%}&0.0\%&54.5\%\\
                
                Llama3-Instruct&70B&70B&16.2\%&-&49.2\%\\
                \midrule
                \shortdsinslite &16B&2.4B&9.2\%&0.0\%&44.4\%\\
                \shortdsins &236B&21B&\textbf{21.0\%}&\textbf{12.7}\%&\textbf{73.7\%}\\
                \bottomrule 
		\end{tabular}
	\caption{\centering Performances of different models on repair benchmarks. We do not evaluate Llama3-Instruct on SWE-Bench as it just supports 8K context length. }        
	\label{table:repair}
        
            \end{small}
\end{table}

Table \ref{table:repair} outlines the performances of different language models on software repair benchmarks, including Defects4J, SWE-Bench, and Aider. Among open-source models, DeepSeek-Coder-Instruct emerges as a standout, achieving the best performance within the open source models. It scores 21\% in Defects4J and 12.7\% in SWE-Bench, closely approaching the results of leading closed-source models and demonstrating significant capability in handling longer code sequences. Notably, DeepSeek-Coder-V2-Instruct achieves the highest score of 73.7\% in Aider, surpassing all other models listed, including closed-source counterparts. This superior performance highlights its efficiency and robustness in automated code repair tasks, positioning DeepSeek-Coder-V2-Instruct as the top open-source model and a formidable competitor to closed-source alternatives in the field.
\subsection{Code Understanding and Reasoning }
To assess the code reasoning capabilities of our models, we utilize the CRUXEval benchmark. This benchmark comprises 800 Python functions paired with corresponding input-output examples. It is divided into two distinct tasks: CRUXEval-I, which requires the large language model (LLM) to predict the output based on the given input, and CRUXEval-O, where the model must predict the input from the known output. This structure challenges the model's ability to understand and reason through Python code in both forward and reverse directions.
\begin{table}[h]
        \centering

		\centering
            \begin{small}
		\resizebox{0.8\linewidth}{!}{\begin{tabular}{lcccccc}
			\toprule
			Model&\#TP&\#AP&CruxEval-I-COT&CruxEval-O-COT\\
               \midrule 
                        \multicolumn{5}{c}{Closed-Source Models} \\ \midrule
                Gemini-1.5-Pro &-&-&67.0\%& 77.5\%\\
                Claude-3-Opus&-&-&73.4\%&82.0\%\\
                GPT-4-1106 &-&-&75.5\%&77.1\%\\
                GPT-4-Turbo-0409 &-& - &{75.7\%}& {82.0\%}\\
                GPT-4o-0513 &-& - &\textbf{77.4\%}& \textbf{88.7\%}\\

               \midrule
                  \multicolumn{5}{c}{Open-Source Models} \\ \midrule
                  
                  Codestral&22B&22B&48.0\% &60.6\% \\
                DS-Coder-Instruct &33B&33B&47.3\%&50.6\%\\
                Llama3-Instruct&70B&70B&61.1\% &64.3\% \\
                \midrule
                \shortdsinslite &16B&2.4B&53.0\%&52.9\%\\
                 \shortdsins &236B&21B&  \textbf{70.0\%}&  \textbf{75.1\%}\\

                \bottomrule 
		\end{tabular}}
	\caption{\centering  Performance of different models on the CruxEval benchmark.}
 \label{table:result-crux}
        \end{small}
\end{table}
Table \ref{table:result-crux} presents the performance of various language models on the CruxEval benchmark, which assesses models on two metrics: CruxEval-I-COT and CruxEval-O-COT. Among the open-source models, DeepSeek-Coder-V2-Instruct stands out significantly. It scores 70.0\% on CruxEval-I-COT and 75.1\% on CruxEval-O-COT, showcasing its superior capability within the open-source domain. However, when compared to larger closed-source models, there is a performance gap. This performance gap may largely be attributed to the fact that DeepSeek-Coder-V2-Instruct operates with only 21 billion activation parameters, which is considerably fewer than those in larger, more advanced closed-source models like GPT-4o. This limitation in model complexity could restrict its learning and problem-solving capacities.
\subsection{Mathematical Reasoning  }
To assess the mathematical reasoning capabilities of DeepSeekCoder-V2, we utilized the popular grade-school benchmark GSM8K \citep{gsm8k}, along with advanced competition-level benchmarks including MATH \citep{MATH}, the American Invitational Mathematics Examination (AIME) 2024 \citep{AIME}, and Math Odyssey \citep{netmindmath}\footnote{The performance of \dscoder on the four mathematical benchmarks was obtained with zero-shot chain-of-thought prompting;
each test question was concatenated with the instruction: "$\backslash$nPlease reason step by step, and put your final answer within $\backslash$boxed\{\}."}.

\begingroup
\setlength{\tabcolsep}{3pt} 
\renewcommand{\arraystretch}{1} 
\begin{table*}[h]
    \centering
\begin{small}

\begin{tabular}{lcccccc} 
\toprule

\multicolumn{1}{l}{\multirow{1}{*}{Model}} & \multicolumn{1}{l}{\multirow{1}{*}{\#TP}}& \multicolumn{1}{l}{\multirow{1}{*}{\#AP}} & GSM8K & MATH & AIME 2024 & Math Odyssey \\
\midrule
\multicolumn{7}{c}{Closed-Source Models} \\ \midrule
Gemini 1.5 Pro & -& -  & 90.8\% & 67.7\% & 2/30 & 45.0\% \\ 
Claude-3-Opus & -& -  & 95.0\% & 60.1\% & 2/30 & 40.6\% \\
GPT-4-1106 & - & - & 91.4\% & 64.3\% & 1/30 & 49.1\% \\
GPT-4-Turbo-0409 & - & - & 93.7\% & 73.4\% & \textbf{3}/\textbf{30} & 46.8\% \\
GPT-4o-0513 & - & - & \textbf{95.8\%} & \textbf{76.6\%} & 2/30 & \textbf{53.2\%} \\ 
 \midrule 
\multicolumn{7}{c}{Open-Source Models} \\ \midrule
Llama3-Instruct & 70B& 70B & 93.0\% & 50.4\% & 1/30 & 27.9\% \\
\shortdsinslite & 16B& 2.4B & 86.4\% & 61.8\% & 0/30 & 44.4\% \\
\shortdsins & 236B& 21B & \textbf{94.9\%} & \textbf{75.7\%} & \textbf{4}/\textbf{30} & \textbf{53.7\%} \\
\bottomrule
\end{tabular}
\caption{
\centering  Performance of different models on the mathematical reasoning. \dscoder-Instruct can achieve 5/30 on AIME 2024 with maj@64. 
   }
    \label{tab:math_res}
\end{small}
\end{table*}
\endgroup

The results, presented in Table \ref{tab:math_res}, were obtained using greedy decoding without the aid of tools or voting techniques, unless otherwise specified.
\dscoder achieved an accuracy of 75.7\% on the MATH benchmark and 53.7\% on Math Odyssey, comparable to the state-of-the-art GPT-4o.
Additionally, \dscoder solves more problems from AIME 2024 than the other models, demonstrating its strong mathematical reasoning capabilities.

\subsection{General Natural Language }
As \dscoder is built upon DeepSeek-V2, it inherits the strong natural language capability, even surpassing DeepSeek-V2 on reasoning-related benchmarks. We compare \dscoder Instruct with DeepSeek-V2 Chat on standard benchmarks, which covers both English and Chinese benchmarks, including  BigBench Hard (BBH) \citep{bbh}, MMLU \citep{mmlu}, ARC \citep{arc}, TriviaQA \citep{joshi-etal-2017-triviaqa}, NaturalQuestions \citep{naturalquestions}, AGIEval \citep{agieval}， CLUEWSC \citep{clue}, C-Eval \citep{ceval}, and CMMLU \citep{cmmlu}.
Besides, we also evaluate the open-ended generation ability of models, including Arena-Hard~\citep{arenahard2024}, AlpacaEval2.0~\citep{alpaca2.0}, MT-Bench~\citep{mtbench}, and Alignbench~\citep{align_bench}.
The evaluation pipeline and metrics are the same as in DeepSeek-V2, where the MMLU are evaluated using OpenAI simple-eval package \url{https://github.com/openai/simple-evals}.

\begin{table}[h]
    \centering
    \footnotesize
    \setlength{\tabcolsep}{5pt}
    \resizebox{\linewidth}{!}{\begin{tabular}{@{}c l c | c  c| c c }
        \toprule
    & \multirow{2}{*}{\centering \textbf{Benchmark (Metric)}} & \multirow{2}{*}{\textbf{\# Shots}} & \textbf{DeepSeek-V2-Lite} & \textbf{\dscoder-Lite} & \textbf{DeepSeek-V2} & \textbf{\dscoder}  \\
    & & & \textbf{Chat} & \textbf{Instruct}  & \textbf{Chat} & \textbf{Instruct}\\
    \midrule
    & \# Active Params & - & 2.4B & 2.4B & 21B & 21B  \\
    & \# Total Params & - & 16B & 16B & 236B & 236B  \\
      & \# Training Tokens & - & 5.7T & 10.2T & 8.1T& 10.2T  \\
    \midrule
    \multirow{7}{*}{English} & BBH (EM) & 3-shot & 48.1 & \bf{61.2} & 79.7 & \bf{83.9}  \\
    & MMLU (Acc.) & 5-shot & 55.7 & \bf{60.1} & 78.1 & \bf{79.2}  \\
    & ARC-Easy (Acc.) & 25-shot & 86.1 & \bf{88.9} & \bf{98.1} & 97.4  \\
    & ARC-Challenge (Acc.) & 25-shot & 73.4 & \bf{77.4} & 92.3 & \bf{92.8}  \\
    & TriviaQA (EM) & 5-shot & \bf{65.2} & 59.5 & \bf{86.7} & 82.3  \\
    & NaturalQuestions (EM) & 5-shot & \bf{35.5} & 30.8 & \bf{53.4} & 47.5  \\
    & AGIEval (Acc.) & 0-shot & \bf{42.8} & 28.7 & \bf{61.4} & 60.0 \\
    \midrule
        \multirow{3}{*}{Chinese} & CLUEWSC (EM) & 5-shot & \bf{80.0} & 76.5 & \bf{89.9} & 85.9  \\
    & C-Eval (Acc.) & 5-shot & 60.1 & \bf{61.6} & 78.0 & \bf{79.4}  \\
    & CMMLU (Acc.) & 5-shot & 62.5 & \bf{62.7} & \bf{81.6} & 80.9  \\
    \midrule
    \multirow{4}{*}{Open-ended} & Arena-Hard& - & 11.40  & \bf{38.10} & 41.60 & \bf{65.00}  \\
    & AlpacaEval 2.0  & - & 16.85 & \bf{17.74} & \bf{38.90} & 36.92  \\
    & MT-Bench & -& 7.37  & \bf{7.81} & \bf{8.97} &  8.77 \\
    & Alignbench & -& 6.02  & \bf{6.83} & \bf{7.91} &  7.84 \\
    \bottomrule
    \end{tabular}
    }
    \caption{
    \centering A Comparison of \dscoder Instruct with DeepSeek-V2 Chat.
    }
    \label{tab:main}
\end{table}
When comparing the performance of 16B models, it is evident that \dscoder-Lite-Instruct outperforms DeepSeek-V2-Lite-Chat in benchmarks like BBH and Arena-Hard. These benchmarks place a high demand on the model's reasoning ability, which \dscoder-Lite-Instruct excels at. However, \dscoder-Lite Instruct falls behind in knowledge-intensive benchmarks like TriviaQA, primarily due to the relatively smaller amount of web data used during pre-training.

Moving on to 236B models, \dscoder Instruct exhibits greater strength in reasoning benchmarks, particularly in Arena-Hard, which comprises a substantial proportion of code, math, and reasoning questions. On the other hand, DeepSeek-V2 Chat demonstrates slightly better results in benchmarks such as  MT-bench \citep{mtbench}, AlpacaEval 2.0 \citep{alpaca2.0}, and AlignBench \citep{align_bench}. This advantage can be attributed to the general-purpose alignment stage of DeepSeek-V2 Chat.

\section{Conclusion}
In this paper, we introduce \dscoder to further advance the field of code intelligence, which is continually pre-trained from DeepSeek-V2 with 6 trillion tokens sourced from a high-quality and multi-source corpus. Through this continued pre-training, we find that \dscoder significantly enhances the model's capabilities in coding and mathematical reasoning while maintaining comparable general language performance to DeepSeek-V2. Compared to DeepSeek-Coder, \dscoder supports a significantly larger number of programming languages, increasing from 86 to 338, and extends the maximum context length from 16K to 128K tokens. Experimental results demonstrate that \dscoder achieves performance comparable to state-of-the-art closed-source models such as GPT-4 Turbo, Claude 3 Opus, and Gemini 1.5 Pro in code and math-specific tasks.

Although \dscoder achieves impressive performance on standard benchmarks, we find that there is still a significant gap in instruction-following capabilities compared to current state-of-the-art models like GPT-4 Turbo. This gap leads to poor performance in complex scenarios and tasks such as those in SWEbench. Therefore, we believe that a code model needs not only strong coding abilities but also exceptional instruction-following capabilities to handle real-world complex programming scenarios. In the future, we will focus more on improving the model's instruction-following capabilities to better handle real-world complex programming scenarios and enhance the productivity of the development process.


\bibliography{main}

\begin{thebibliography}{43}
\providecommand{\natexlab}[1]{#1}
\providecommand{\url}[1]{\texttt{#1}}
\expandafter\ifx\csname urlstyle\endcsname\relax
  \providecommand{\doi}[1]{doi: #1}\else
  \providecommand{\doi}{doi: \begingroup \urlstyle{rm}\Url}\fi

\bibitem[Allal et~al.(2023)Allal, Li, Kocetkov, Mou, Akiki, Ferrandis, Muennighoff, Mishra, Gu, Dey, et~al.]{allal2023santacoder}
L.~B. Allal, R.~Li, D.~Kocetkov, C.~Mou, C.~Akiki, C.~M. Ferrandis, N.~Muennighoff, M.~Mishra, A.~Gu, M.~Dey, et~al.
\newblock Santacoder: don't reach for the stars!
\newblock \emph{arXiv preprint arXiv:2301.03988}, 2023.

\bibitem[Anthropic(2024)]{anthropic2024claude}
A.~Anthropic.
\newblock The claude 3 model family: Opus, sonnet, haiku.
\newblock \emph{Claude-3 Model Card}, 2024.

\bibitem[Austin et~al.(2021{\natexlab{a}})Austin, Odena, Nye, Bosma, Michalewski, Dohan, Jiang, Cai, Terry, Le, and Sutton]{austin2021program}
J.~Austin, A.~Odena, M.~Nye, M.~Bosma, H.~Michalewski, D.~Dohan, E.~Jiang, C.~Cai, M.~Terry, Q.~Le, and C.~Sutton.
\newblock Program synthesis with large language models, 2021{\natexlab{a}}.

\bibitem[Austin et~al.(2021{\natexlab{b}})Austin, Odena, Nye, Bosma, Michalewski, Dohan, Jiang, Cai, Terry, Le, et~al.]{mbpp}
J.~Austin, A.~Odena, M.~Nye, M.~Bosma, H.~Michalewski, D.~Dohan, E.~Jiang, C.~Cai, M.~Terry, Q.~Le, et~al.
\newblock Program synthesis with large language models.
\newblock \emph{arXiv preprint arXiv:2108.07732}, 2021{\natexlab{b}}.

\bibitem[Bavarian et~al.(2022)Bavarian, Jun, Tezak, Schulman, McLeavey, Tworek, and Chen]{bavarian2022efficient}
M.~Bavarian, H.~Jun, N.~Tezak, J.~Schulman, C.~McLeavey, J.~Tworek, and M.~Chen.
\newblock Efficient training of language models to fill in the middle.
\newblock \emph{arXiv preprint arXiv:2207.14255}, 2022.

\bibitem[Chen et~al.(2021)Chen, Tworek, Jun, Yuan, Pinto, Kaplan, Edwards, Burda, Joseph, Brockman, et~al.]{chen2021evaluating}
M.~Chen, J.~Tworek, H.~Jun, Q.~Yuan, H.~P. d.~O. Pinto, J.~Kaplan, H.~Edwards, Y.~Burda, N.~Joseph, G.~Brockman, et~al.
\newblock Evaluating large language models trained on code.
\newblock \emph{arXiv preprint arXiv:2107.03374}, 2021.

\bibitem[Clark et~al.(2018)Clark, Cowhey, Etzioni, Khot, Sabharwal, Schoenick, and Tafjord]{arc}
P.~Clark, I.~Cowhey, O.~Etzioni, T.~Khot, A.~Sabharwal, C.~Schoenick, and O.~Tafjord.
\newblock Think you have solved question answering? try arc, the {AI2} reasoning challenge.
\newblock \emph{CoRR}, abs/1803.05457, 2018.
\newblock URL \url{http://arxiv.org/abs/1803.05457}.

\bibitem[Cobbe et~al.(2021)Cobbe, Kosaraju, Bavarian, Chen, Jun, Kaiser, Plappert, Tworek, Hilton, Nakano, et~al.]{gsm8k}
K.~Cobbe, V.~Kosaraju, M.~Bavarian, M.~Chen, H.~Jun, L.~Kaiser, M.~Plappert, J.~Tworek, J.~Hilton, R.~Nakano, et~al.
\newblock Training verifiers to solve math word problems.
\newblock \emph{arXiv preprint arXiv:2110.14168}, 2021.

\bibitem[DeepSeek-AI(2024)]{deepseekai2024deepseekv2}
DeepSeek-AI.
\newblock Deepseek-v2: A strong, economical, and efficient mixture-of-experts language model, 2024.

\bibitem[Dubois et~al.(2024)Dubois, Galambosi, Liang, and Hashimoto]{alpaca2.0}
Y.~Dubois, B.~Galambosi, P.~Liang, and T.~B. Hashimoto.
\newblock Length-controlled alpacaeval: A simple way to debias automatic evaluators.
\newblock \emph{arXiv preprint arXiv:2404.04475}, 2024.

\bibitem[Guo et~al.(2024)Guo, Zhu, Yang, Xie, Dong, Zhang, Chen, Bi, Wu, Li, et~al.]{guo2024deepseek}
D.~Guo, Q.~Zhu, D.~Yang, Z.~Xie, K.~Dong, W.~Zhang, G.~Chen, X.~Bi, Y.~Wu, Y.~Li, et~al.
\newblock Deepseek-coder: When the large language model meets programming--the rise of code intelligence.
\newblock \emph{arXiv preprint arXiv:2401.14196}, 2024.

\bibitem[Hendrycks et~al.(2020)Hendrycks, Burns, Basart, Zou, Mazeika, Song, and Steinhardt]{mmlu}
D.~Hendrycks, C.~Burns, S.~Basart, A.~Zou, M.~Mazeika, D.~Song, and J.~Steinhardt.
\newblock Measuring massive multitask language understanding.
\newblock \emph{arXiv preprint arXiv:2009.03300}, 2020.

\bibitem[Hendrycks et~al.(2021)Hendrycks, Burns, Kadavath, Arora, Basart, Tang, Song, and Steinhardt]{MATH}
D.~Hendrycks, C.~Burns, S.~Kadavath, A.~Arora, S.~Basart, E.~Tang, D.~Song, and J.~Steinhardt.
\newblock Measuring mathematical problem solving with the math dataset.
\newblock \emph{arXiv preprint arXiv:2103.03874}, 2021.

\bibitem[Huang et~al.(2023)Huang, Bai, Zhu, Zhang, Zhang, Su, Liu, Lv, Zhang, Lei, et~al.]{ceval}
Y.~Huang, Y.~Bai, Z.~Zhu, J.~Zhang, J.~Zhang, T.~Su, J.~Liu, C.~Lv, Y.~Zhang, J.~Lei, et~al.
\newblock {C-Eval}: A multi-level multi-discipline chinese evaluation suite for foundation models.
\newblock \emph{arXiv preprint arXiv:2305.08322}, 2023.

\bibitem[Jain et~al.(2024)Jain, Han, Gu, Li, Yan, Zhang, Wang, Solar-Lezama, Sen, and Stoica]{jain2024livecodebench}
N.~Jain, K.~Han, A.~Gu, W.-D. Li, F.~Yan, T.~Zhang, S.~Wang, A.~Solar-Lezama, K.~Sen, and I.~Stoica.
\newblock Livecodebench: Holistic and contamination free evaluation of large language models for code, 2024.

\bibitem[Jimenez et~al.(2023)Jimenez, Yang, Wettig, Yao, Pei, Press, and Narasimhan]{jimenez2023swe}
C.~E. Jimenez, J.~Yang, A.~Wettig, S.~Yao, K.~Pei, O.~Press, and K.~Narasimhan.
\newblock Swe-bench: Can language models resolve real-world github issues?
\newblock \emph{arXiv preprint arXiv:2310.06770}, 2023.

\bibitem[Joshi et~al.(2017)Joshi, Choi, Weld, and Zettlemoyer]{joshi-etal-2017-triviaqa}
M.~Joshi, E.~Choi, D.~Weld, and L.~Zettlemoyer.
\newblock {T}rivia{QA}: A large scale distantly supervised challenge dataset for reading comprehension.
\newblock In R.~Barzilay and M.-Y. Kan, editors, \emph{Proceedings of the 55th Annual Meeting of the Association for Computational Linguistics (Volume 1: Long Papers)}, pages 1601--1611, Vancouver, Canada, July 2017. Association for Computational Linguistics.
\newblock \doi{10.18653/v1/P17-1147}.
\newblock URL \url{https://aclanthology.org/P17-1147}.

\bibitem[Joulin et~al.(2016)Joulin, Grave, Bojanowski, Douze, J{\'e}gou, and Mikolov]{joulin2016fasttext}
A.~Joulin, E.~Grave, P.~Bojanowski, M.~Douze, H.~J{\'e}gou, and T.~Mikolov.
\newblock Fasttext. zip: Compressing text classification models.
\newblock \emph{arXiv preprint arXiv:1612.03651}, 2016.

\bibitem[Kocetkov et~al.(2022)Kocetkov, Li, Jia, Mou, Jernite, Mitchell, Ferrandis, Hughes, Wolf, Bahdanau, et~al.]{kocetkov2022stack}
D.~Kocetkov, R.~Li, L.~Jia, C.~Mou, Y.~Jernite, M.~Mitchell, C.~M. Ferrandis, S.~Hughes, T.~Wolf, D.~Bahdanau, et~al.
\newblock The stack: 3 tb of permissively licensed source code.
\newblock \emph{Transactions on Machine Learning Research}, 2022.

\bibitem[Kwiatkowski et~al.(2019)Kwiatkowski, Palomaki, Redfield, Collins, Parikh, Alberti, Epstein, Polosukhin, Devlin, Lee, Toutanova, Jones, Kelcey, Chang, Dai, Uszkoreit, Le, and Petrov]{naturalquestions}
T.~Kwiatkowski, J.~Palomaki, O.~Redfield, M.~Collins, A.~P. Parikh, C.~Alberti, D.~Epstein, I.~Polosukhin, J.~Devlin, K.~Lee, K.~Toutanova, L.~Jones, M.~Kelcey, M.~Chang, A.~M. Dai, J.~Uszkoreit, Q.~Le, and S.~Petrov.
\newblock Natural questions: a benchmark for question answering research.
\newblock \emph{Trans. Assoc. Comput. Linguistics}, 7:\penalty0 452--466, 2019.
\newblock \doi{10.1162/tacl\_a\_00276}.
\newblock URL \url{https://doi.org/10.1162/tacl\_a\_00276}.

\bibitem[Li et~al.(2023{\natexlab{a}})Li, Zhang, Koto, Yang, Zhao, Gong, Duan, and Baldwin]{cmmlu}
H.~Li, Y.~Zhang, F.~Koto, Y.~Yang, H.~Zhao, Y.~Gong, N.~Duan, and T.~Baldwin.
\newblock {CMMLU}: Measuring massive multitask language understanding in {Chinese}.
\newblock \emph{arXiv preprint arXiv:2306.09212}, 2023{\natexlab{a}}.

\bibitem[Li et~al.(2023{\natexlab{b}})Li, Allal, Zi, Muennighoff, Kocetkov, Mou, Marone, Akiki, Li, Chim, et~al.]{li2023starcoder}
R.~Li, L.~B. Allal, Y.~Zi, N.~Muennighoff, D.~Kocetkov, C.~Mou, M.~Marone, C.~Akiki, J.~Li, J.~Chim, et~al.
\newblock Starcoder: may the source be with you!
\newblock \emph{arXiv preprint arXiv:2305.06161}, 2023{\natexlab{b}}.

\bibitem[Li et~al.(2024)Li, Chiang, Frick, Dunlap, Zhu, Gonzalez, and Stoica]{arenahard2024}
T.~Li, W.-L. Chiang, E.~Frick, L.~Dunlap, B.~Zhu, J.~E. Gonzalez, and I.~Stoica.
\newblock From live data to high-quality benchmarks: The arena-hard pipeline, April 2024.
\newblock URL \url{https://lmsys.org/blog/2024-04-19-arena-hard/}.

\bibitem[Liu et~al.(2023{\natexlab{a}})Liu, Xia, Wang, and Zhang]{evalplus}
J.~Liu, C.~S. Xia, Y.~Wang, and L.~Zhang.
\newblock Is your code generated by chat{GPT} really correct? rigorous evaluation of large language models for code generation.
\newblock In \emph{Thirty-seventh Conference on Neural Information Processing Systems}, 2023{\natexlab{a}}.
\newblock URL \url{https://openreview.net/forum?id=1qvx610Cu7}.

\bibitem[Liu et~al.(2023{\natexlab{b}})Liu, Xu, and McAuley]{liu2023repobench}
T.~Liu, C.~Xu, and J.~McAuley.
\newblock Repobench: Benchmarking repository-level code auto-completion systems.
\newblock In \emph{The Twelfth International Conference on Learning Representations}, 2023{\natexlab{b}}.

\bibitem[Liu et~al.(2023{\natexlab{c}})Liu, Lei, Wang, Huang, Feng, Wen, Cheng, Ke, Xu, Tam, Zhang, Sun, Wang, Zhang, Huang, Dong, and Tang]{align_bench}
X.~Liu, X.~Lei, S.~Wang, Y.~Huang, Z.~Feng, B.~Wen, J.~Cheng, P.~Ke, Y.~Xu, W.~L. Tam, X.~Zhang, L.~Sun, H.~Wang, J.~Zhang, M.~Huang, Y.~Dong, and J.~Tang.
\newblock Alignbench: Benchmarking chinese alignment of large language models.
\newblock \emph{CoRR}, abs/2311.18743, 2023{\natexlab{c}}.
\newblock \doi{10.48550/ARXIV.2311.18743}.
\newblock URL \url{https://doi.org/10.48550/arXiv.2311.18743}.

\bibitem[Loshchilov and Hutter(2019)]{loshchilov2019decoupled}
I.~Loshchilov and F.~Hutter.
\newblock Decoupled weight decay regularization, 2019.

\bibitem[Lozhkov et~al.(2024)Lozhkov, Li, Allal, Cassano, Lamy-Poirier, Tazi, Tang, Pykhtar, Liu, Wei, et~al.]{lozhkov2024starcoder}
A.~Lozhkov, R.~Li, L.~B. Allal, F.~Cassano, J.~Lamy-Poirier, N.~Tazi, A.~Tang, D.~Pykhtar, J.~Liu, Y.~Wei, et~al.
\newblock Starcoder 2 and the stack v2: The next generation.
\newblock \emph{arXiv preprint arXiv:2402.19173}, 2024.

\bibitem[MAA(2024)]{AIME}
MAA.
\newblock American invitational mathematics examination - aime.
\newblock \emph{American Invitational Mathematics Examination - AIME 2024}, 2024.
\newblock URL \url{https://maa.org/math-competitions/american-invitational-mathematics-examination-aime}.

\bibitem[Meta(2024)]{meta2024llama3}
Meta.
\newblock Introducing meta llama 3: The most capable openly available llm to date.
\newblock \url{https://ai.meta.com/blog/meta-llama-3/}, April 2024.

\bibitem[MistralAI(2024)]{mistral2024codestral}
MistralAI.
\newblock Codestral.
\newblock \url{https://mistral.ai/news/codestral/}, 2024.
\newblock Accessed: 2024-05-29.

\bibitem[Netmind.AI(2024)]{netmindmath}
Netmind.AI.
\newblock Odyssey-math.
\newblock \url{https://github.com/protagolabs/odyssey-math/tree/main}, 2024.
\newblock Accessed: April 22, 2024.

\bibitem[OpenAI(2023)]{openai2023gpt4}
OpenAI.
\newblock Gpt-4 technical report, 2023.

\bibitem[Peng et~al.(2023)Peng, Quesnelle, Fan, and Shippole]{peng2023yarn}
B.~Peng, J.~Quesnelle, H.~Fan, and E.~Shippole.
\newblock Yarn: Efficient context window extension of large language models.
\newblock \emph{arXiv preprint arXiv:2309.00071}, 2023.

\bibitem[Reid et~al.(2024)Reid, Savinov, Teplyashin, Lepikhin, Lillicrap, Alayrac, Soricut, Lazaridou, Firat, Schrittwieser, et~al.]{reid2024gemini}
M.~Reid, N.~Savinov, D.~Teplyashin, D.~Lepikhin, T.~Lillicrap, J.-b. Alayrac, R.~Soricut, A.~Lazaridou, O.~Firat, J.~Schrittwieser, et~al.
\newblock Gemini 1.5: Unlocking multimodal understanding across millions of tokens of context.
\newblock \emph{arXiv preprint arXiv:2403.05530}, 2024.

\bibitem[Roziere et~al.(2023)Roziere, Gehring, Gloeckle, Sootla, Gat, Tan, Adi, Liu, Remez, Rapin, et~al.]{roziere2023code}
B.~Roziere, J.~Gehring, F.~Gloeckle, S.~Sootla, I.~Gat, X.~E. Tan, Y.~Adi, J.~Liu, T.~Remez, J.~Rapin, et~al.
\newblock Code llama: Open foundation models for code.
\newblock \emph{arXiv preprint arXiv:2308.12950}, 2023.

\bibitem[Shao et~al.(2024)Shao, Wang, Zhu, Xu, Song, Zhang, Li, Wu, and Guo]{shao2024deepseekmath}
Z.~Shao, P.~Wang, Q.~Zhu, R.~Xu, J.~Song, M.~Zhang, Y.~Li, Y.~Wu, and D.~Guo.
\newblock Deepseekmath: Pushing the limits of mathematical reasoning in open language models.
\newblock \emph{arXiv preprint arXiv:2402.03300}, 2024.

\bibitem[Shi et~al.(2024)Shi, Tang, Narasimhan, and Yao]{usaco}
Q.~Shi, M.~Tang, K.~Narasimhan, and S.~Yao.
\newblock Can language models solve olympiad programming?
\newblock \emph{arXiv preprint arXiv:2404.10952}, 2024.

\bibitem[Suzgun et~al.(2022)Suzgun, Scales, Sch{\"a}rli, Gehrmann, Tay, Chung, Chowdhery, Le, Chi, Zhou, et~al.]{bbh}
M.~Suzgun, N.~Scales, N.~Sch{\"a}rli, S.~Gehrmann, Y.~Tay, H.~W. Chung, A.~Chowdhery, Q.~V. Le, E.~H. Chi, D.~Zhou, et~al.
\newblock Challenging big-bench tasks and whether chain-of-thought can solve them.
\newblock \emph{arXiv preprint arXiv:2210.09261}, 2022.

\bibitem[Touvron et~al.(2023)Touvron, Martin, Stone, Albert, Almahairi, Babaei, Bashlykov, Batra, Bhargava, Bhosale, et~al.]{touvron2023llama}
H.~Touvron, L.~Martin, K.~Stone, P.~Albert, A.~Almahairi, Y.~Babaei, N.~Bashlykov, S.~Batra, P.~Bhargava, S.~Bhosale, et~al.
\newblock Llama 2: Open foundation and fine-tuned chat models.
\newblock \emph{arXiv preprint arXiv:2307.09288}, 2023.

\bibitem[Xu et~al.(2020)Xu, Hu, Zhang, Li, Cao, Li, Xu, Sun, Yu, Yu, Tian, Dong, Liu, Shi, Cui, Li, Zeng, Wang, Xie, Li, Patterson, Tian, Zhang, Zhou, Liu, Zhao, Zhao, Yue, Zhang, Yang, Richardson, and Lan]{clue}
L.~Xu, H.~Hu, X.~Zhang, L.~Li, C.~Cao, Y.~Li, Y.~Xu, K.~Sun, D.~Yu, C.~Yu, Y.~Tian, Q.~Dong, W.~Liu, B.~Shi, Y.~Cui, J.~Li, J.~Zeng, R.~Wang, W.~Xie, Y.~Li, Y.~Patterson, Z.~Tian, Y.~Zhang, H.~Zhou, S.~Liu, Z.~Zhao, Q.~Zhao, C.~Yue, X.~Zhang, Z.~Yang, K.~Richardson, and Z.~Lan.
\newblock {CLUE:} {A} chinese language understanding evaluation benchmark.
\newblock In D.~Scott, N.~Bel, and C.~Zong, editors, \emph{Proceedings of the 28th International Conference on Computational Linguistics, {COLING} 2020, Barcelona, Spain (Online), December 8-13, 2020}, pages 4762--4772. International Committee on Computational Linguistics, 2020.
\newblock \doi{10.18653/V1/2020.COLING-MAIN.419}.
\newblock URL \url{https://doi.org/10.18653/v1/2020.coling-main.419}.

\bibitem[Zheng et~al.(2023)Zheng, Chiang, Sheng, Zhuang, Wu, Zhuang, Lin, Li, Li, Xing, Zhang, Gonzalez, and Stoica]{mtbench}
L.~Zheng, W.-L. Chiang, Y.~Sheng, S.~Zhuang, Z.~Wu, Y.~Zhuang, Z.~Lin, Z.~Li, D.~Li, E.~P. Xing, H.~Zhang, J.~E. Gonzalez, and I.~Stoica.
\newblock Judging llm-as-a-judge with mt-bench and chatbot arena, 2023.

\bibitem[Zhong et~al.(2023)Zhong, Cui, Guo, Liang, Lu, Wang, Saied, Chen, and Duan]{agieval}
W.~Zhong, R.~Cui, Y.~Guo, Y.~Liang, S.~Lu, Y.~Wang, A.~Saied, W.~Chen, and N.~Duan.
\newblock {AGIEval}: {A} human-centric benchmark for evaluating foundation models.
\newblock \emph{CoRR}, abs/2304.06364, 2023.
\newblock \doi{10.48550/arXiv.2304.06364}.
\newblock URL \url{https://doi.org/10.48550/arXiv.2304.06364}.

\end{thebibliography}

\newpage
\appendix

\section{Supported Programming Languages}
\label{sec:supported_pl}
ABAP, ActionScript, Ada, Agda, AGS Script, Alloy, AmbientTalk, AMD GPU, AMPL, ANSYS Parametric Design Language, ANTLR, Apache Configuration, APL, AppleScript, Arc, Arduino, ASP, AspectJ, Assembly, Asymptote, Augeas, AutoHotkey, AutoIt, AWK, BC, Berry, BitBake, BlitzBasic, BlitzMax, Bluespec, BNF, Boo, Boogie, Brainfuck, BrightScript, Bro, BST, C, C\#, C2HS Haskell, CADL, CapDL, Ceylon, Chapel, ChucK, Cirru, Click, Clojure, CMake, COBOL, COBOLFree, CoffeeScript, ColdFusion CFC, Common Lisp, C++, Crystal, Csound, Csound Score, CSS, CUDA, Cypher, Cython, Darcs Patch, Dart, DASM16, Debian Control File, DeviceTree, Diff, DM, Docker, Dockerfile, Dylan, EBNF, eC, Eiffel, Elixir, Elm, ELPi, Emacs Lisp, EmberScript, Erlang, Execline, F\#, Factor, Fancy, Fantom, Felix, Fennel, Fish, Flux, Fortran, Fortran Fixed Form, FoxPro, FreeFem, FreeMarker, F*, Futhark, G-Code, GAP, GAS, GDScript, Genshi, Gentoo Ebuild, Gentoo Eclass, Gettext Catalog, GLSL, Glyph, Gnuplot, Go, Gosu, Grace, Gradle, Grammatical Framework, GraphQL, Graphviz DOT, Groff, Groovy, Groovy Server Pages, GSQL, Handlebars, Haskell, Haxe, HCL, HLSL, HTML, HTML Django, HTML ERB, HTML PHP, HTTP, Hy, Idris, IGOR Pro, Inform 6 Template, Inno Setup, Io, Isabelle, J, Jade, JAGS, Jasmin, Java, Java Server Pages, JavaScript, JavaScript MozPreproc, JCL, JFlex, JSON, JSONiq, JSX, Julia, Jupyter Notebook, K, Kconfig, Koka, Kotlin, KRL, Lean, Less, Lex, LFE, Lighttpd Configuration File, LilyPond, Limbo, Linker Script, Liquid, Literate Agda, Literate CoffeeScript, LLVM, Logtalk, LSL, Lua, M4, Makefile, Mako, Mason, MATLAB, Maxima, Meson, Metal, MiniScript, Mirah, Mizar, Modelica, Modula-2, Monkey, MooCode, MoonScript, Mosel, MQL, MUF, MuPAD, NASM, NCL, NetLinx, Nginx Configuration File, Nimrod, Ninja, Nit, Nix, NSIS, Nu, NuSMV, Objdump, Objective-C, Objective-C++, OCaml, Octave, Odin, OMG Interface Definition Language, ooc, Opa, OpenCL, OpenEdge ABL, OpenSCAD, Ox, Oz, Papyrus, Parrot Internal Representation, Pascal, PAWN, PEG, Perl, Perl 6, PHP, Pike, PkgConfig, POD, Pony, POV-Ray, PowerShell, Praat, Processing, Propeller Spin, Protocol Buffer, Pug, Puppet, PureBasic, PureScript, Python, Q, QML, QVTO, R, Racket, Ragel in Ruby Host, RAML, RConsole, Rd, REALbasic, ReasonML, Red, RenderScript, Ren'Py, REXX, RHTML, Ride, Robot Framework, Rouge, Ruby, Rust, S, Sage, SARL, SAS, Sass, Scala, Scheme, Scilab, SCSS, Self, Shell, ShExC, Sieve, Silver, Singularity, Slim, Smali, Smarty, Smithy, SMT, Solidity, SourcePawn, SPARQL, SQF, SQL, Squirrel, Stan, Standard ML, Stata, Stylus, SuperCollider, Swift, SWIG, SystemVerilog, Tcl, Tcsh, Tea, Terminfo, TeX, Thrift, Transact-SQL, Treetop, Turing, Twig, TypeScript, TypoScript, Unity3D Asset, Uno, UnrealScript, UrWeb, USD, Vala, VBScript, VCL, Velocity, Verilog, VHDL, VimL, Visual Basic, Vue, WebAssembly, Web IDL, Whiley, X10, XBase, XC, XML, XML Lasso, XQuery, XS, XSLT, Xtend, Xtlang, YANG, Zeek, Zephir, Zig, Zimpl

\setcounter{figure}{0}
\makeatletter 
\renewcommand{\thefigure}{A\@arabic\c@figure}
\makeatother

\setcounter{table}{0}
\makeatletter 
\renewcommand{\thetable}{A\@arabic\c@table}
\makeatother

\end{CJK*}
\end{document}